\documentclass{rrparticle}

\usepackage{graphicx}
\usepackage[usenames]{color}
\usepackage{ulem} 
\title{Fractional exclusion statistics in non-homogeneous interacting particle systems}

\author[1,2]{G. A. Nemnes}
\author[2]{D. V. Anghel}

\affil[1]{University of Bucharest, Faculty of Physics, ``Materials and Devices for Electronics and Optoelectronics'' Research Center, P.O. Box MG-11, 077125 M\u agurele-Ilfov, Romania.}
\affil[2]{Horia Hulubei National Institute for Physics and Nuclear Engineering, P.O. Box MG-6, 077126 M\u agurele-Ilfov, Romania.}

\newcommand{\rr}{{\bf r}}

\newcommand{\cZ}{{\mathcal Z}}

\newcommand{\emin}{{\epsilon_{\rm min}}}
\newcommand{\emax}{{\epsilon_{\rm max}}}
\newcommand{\temin}{{\tilde\epsilon_{\rm min}}}
\newcommand{\temax}{{\tilde\epsilon_{\rm max}}}

\newcommand{\kB}{{k_{\rm B}}}
\newcommand{\WB}{{W^{(-)}}}
\newcommand{\WF}{{W^{(+)}}}

\newcommand{\rmW}{{\rm W}}

\begin{document}

\maketitle

\begin{abstract}
We develop a model based on the fractional exclusion statistics (FES) applicable to non-homogeneous interacting particle systems. Here the species represent elementary volumes in an $(s+1)$-dimensional space, formed by the direct product between the $s$-dimensional space of positions and the quasiparticle energy axis. The model is particularly suitable for systems with localized states. 
We prove the feasibility of our method by applying it to systems of different degrees of complexities.
We first apply the formalism on simpler systems, formed of two sub-systems, and present numerical and
analytical thermodynamic calculations, pointing out the quasiparticle population inversion
and maxima in the heat capacity, in contrast to systems with only diagonal (direct) FES parameters.
Further we investigate larger, non-homogeneous systems with repulsive screened Coulomb interactions, 
indicating accumulation and depletion effects at the interfaces.
Finally, we consider systems with several degrees of disorder, which are
prototypical for models with glassy behavior. 
We find that the disorder produces a spatial segregation of quasiparticles
at low energies which significantly affects the heat capacity and the entropy of the system.
\end{abstract}

\section{Introduction}
\label{intro}

The concept of fractional exclusion statistics (FES), which is a generalization of the Pauli exclusion principle, was introduced by Haldane in Ref. \cite{PhysRevLett.67.937.1991.Haldane} and the statistical mechanics of FES systems was formulated by several authors, employing different methods \cite{PhysRevD.45.4706.1992.Ramanathan,PhysRevLett.72.600.1994.Veigy,PhysRevLett.73.922.1994.Wu,PhysRevLett.73.2150.1994.Isakov,PhysRevLett.73.3331.1994.Murthy,FES_intro2013.Murthy}.
Other generalizations of the Bose and Fermi statistics include the {\it Gentile's} statistics \cite{NuovoCim.17.493.1940.Gentile,NuovoCim.19.106.1942.Gentile}, obtained 
by fixing the maximum occupation number of a single-particle state,
{\it anyonic statistics} \cite{NuovoCim.B37.1.1977.Leinaas,NuovoCim.A4.19.1978.Leinaas,FortschrPhys.28.579.1980.Leinaas,PhysRevLett.49.957.1982.Wilczek,JMathPhys.45.3770.2004.Goldin}, which is connected to the braid group, and 
{\it q-deformed statistics}, concerning systems of particles in arbitrary dimensions which satisfy quon algebras \cite{Symmetries_in_Science_X.1998.Daoud}.

The FES was applied to quasiparticle excitations at the lowest Landau level in the fractional quantum Hall effect, spinon excitations in a spin-$\frac{1}{2}$ quantum antiferomagnet \cite{PhysRevLett.67.937.1991.Haldane,PhysRevLett.73.922.1994.Wu,IntJModPhysA12.1895.1997.Isakov}, Bose and Fermi systems described in the thermodynamic Bethe ansatz \cite{PhysRevLett.73.2150.1994.Isakov,NewDevIntSys.1995.Bernard,PhysRevE.75.61120.2007.Potter,PhysRevE.76.61112.2007.Potter}, excitations \cite{PhysRevLett.67.937.1991.Haldane} or motifs of spins \cite{PhysRevE.84.021136.2011.Liu,PhysRevE.85.011144.2012.Liu} in spin chains, elementary volumes obtained by coarse-graining in the phase-space of a system \cite{PhysRevLett.73.2150.1994.Isakov,PhysRevB.56.4422.1997.Sutherland,PhysRevLett.85.2781.2000.Iguchi,JPhysA.40.F1013.2007.Anghel,PhysLettA.372.5745.2008.Anghel,PhysLettA.376.892.2012.Anghel,EPL.90.10006.2010.Anghel},
interacting particles described in the mean-field approximation \cite{PhysRevLett.74.3912.1995.Sen,JPhysB33.3895.2000.Bhaduri,PhysRevLett.86.2930.2001.Hansson,PhysRevE.88.042150.2013.Anghel}, etc.
Generic FES systems in different numbers of dimensions or interacting with external fields have also been studied for example in Refs.
\cite{PhysRevE.78.021127.2008.Mirza,PhysRevE.80.011132.2009.Mirza,PhysRevE.82.031137.2010.Mirza,PhysRevE.51.3729.1995.Huang,PRB.53.15842.1996.Huang,JPhysB.43.055302.2010.Qin,PhysRevE.83.021111.2011.Qin,PhysLettA.376.1191.2012.Qin,CommunTeorPhys.58.573.2012.Qin}.
Correlations in FES systems have been calculated in \cite{PhysRevE.76.061123.2007.Pellegrino}.
In Refs. \cite{PhysRevLett.74.3912.1995.Sen,PhysLettA212.299.1996.Isakov} it was shown that systems of constant density of states with equal and diagonal FES parameters have the same thermodynamic properties under canonical conditions. This property extends the Bose-Fermi thermodynamic equivalence in 2D systems discovered long time ago \cite{ProcCambrPhilos42.272.1946.Auluc,PhysRev.135.A1515.1964.May,PhysRevE.55.1518.1997.Lee}, to the FES systems \cite{JPA35.7255.2002.Anghel}.

A stochastic method for the simulation of the time evolution of FES systems was introduced in Ref. \cite{JStatMech.P09011.2010.Nemnes} as a generalization of a similar method used for Bose and Fermi systems \cite{JStatMech.2009.P02021.2009.Guastella} whereas the relatively recent experimental realization of the Fermi degeneracy in cold atomic gases have renewed the interest in the theoretical investigation of non-ideal Fermi systems at low temperatures and their interpretation as ideal FES systems 
%
%
\cite{JPhysB42.235302.2009.Bhaduri,JPhysB.43.055302.2010.Qin,PhysRevE.83.021111.2011.Qin,JPhysA.45.315302.2012.vanZyl,JPhysA.46.045001.2013.MacDonald}.

The FES formalism was amended to include the change of the FES parameters at the change of the particle species \cite{JPhysA.40.F1013.2007.Anghel,EPL.87.60009.2009.Anghel} and this allows the implementation of FES as a general method to describe interacting particle systems as ideal gases of quasiparticles
\cite{PhysLettA.372.5745.2008.Anghel,PhysLettA.376.892.2012.Anghel,JPhysConfSer.410.012120.2013.Nemnes,PhysRevE.88.042150.2013.Anghel,PhysScr.2012.014079.2012.Anghel,PhysRevE.84.021136.2011.Liu,PhysRevE.85.011144.2012.Liu}.
Moreover, FES was applied also to systems of classical particles in Refs. \cite{JStatMech.P04018.2013.Gundlach,arXiv:1312.5854.Moore}.

In this paper we present the procedure of describing systems of particles with long-range interaction as ideal FES systems. The method was introduced briefly in Refs. \cite{JPhysConfSer.410.012120.2013.Nemnes,PhysRevE.88.042150.2013.Anghel} and we apply it here to systems consisting of randomly and (in general) non-uniformly distributed localized states which can be occupied by particles.
The particle-particle interaction potential depends only on the distance $r$ between the particles, like for example the screened Coulomb interaction, $V(r)\propto\exp(-r/\lambda)/r^\gamma$.
We define the quasiparticle energies and the fractional exclusion statistics parameters. Using these we write the FES equations for equilibrium particle distribution. In Ref. \cite{PhysRevE.88.042150.2013.Anghel} it was shown that the solution provided by the FES equations is equivalent to the standard solution by Landau's Fermi liquid theory (FLT). 
We show here the feasibility of the FES method by solving numerically the FES equations for a variety of one-dimensional systems, with homogeneous and non-homogeneous particle distributions.

This method may find applications in glassy systems such as the Coulomb glasses, systems of bosons trapped in optical lattices, but also to mesoscopic transport in an already implemented Monte Carlo framework \cite{JStatMech.P09011.2010.Nemnes}.

Another FES method for the description of one-dimensional systems of particles which occupy localized states, 
was proposed in \cite{arXiv:1311.7574.Bakhti}. It is not our purpose to compare the two methods here, but they may be complementary to each other. 
Our method may describe systems of particles with long range interaction in any number of dimensions, with non-homogeneous sites distributions and different energies per site, but assumes a large number of particles and sites within the range of the particle-particle interaction (see also Refs. \cite{JPhysConfSer.410.012120.2013.Nemnes,PhysRevE.88.042150.2013.Anghel}).
On the other hand, the method of Ref. \cite{arXiv:1311.7574.Bakhti} describes one-dimensional homogeneous lattice gases of the same energy per site, on the basis of statistically interacting vacancy particles. The method is 
accurate for any range of the particle-particle interaction and is especially suitable for extracting
the distributions of spaces between individual particles.

The structure of the paper is as follows.
In the following subsection we introduce briefly the notations and the basic concepts of FES. 
In Section \ref{model} we introduce our model, in which species are elementary volumes in the $(s+1)$-dimensional space formed by the direct product between the $s$-dimensional space of positions, $\Omega$, and the energy axis, $\epsilon$ or $\tilde\epsilon$. These species are related by FES parameters, which we calculate. 
We prove the feasibility of the method by applying it to a few physically relevant systems of different complexities.
First we apply it to homogeneous systems of particles interacting by repulsive (screened) Coulomb potentials.
Then, in Section \ref{applications}, we apply our formalism on 
a few test cases with reduced number of species, which are analytically
tractable in order to capture essential features of interacting
inhomogeneous systems. Next, the accumulation and depletion of particles
subject to screened Coulomb interactions is investigated 
in larger, non-uniform systems, and interface phenomena are emphasized.
Finally, systems with several degrees of disorder are considered and the 
spatial segregation of quasiparticles is pointed out 
together with its consequences in the thermodynamic behavior.

\subsection{Basic definitions}

A FES system consists of a countable set of species, indexed by $i,j=0,1,2,\ldots$. Each species contains a finite number of single-particle states and particles, denoted by $G_i$ and $N_i$, respectively. The number of states in the species depend on the number of particles. For small variations of the number of particles around some reference distribution, $\{N_i\}_{i=0,1,...}$, the number of states changes by 
\begin{equation}
 \delta G_i=-\sum_j \alpha_{ij}\delta N_j, \label{alpha_def}
\end{equation}
where by $\delta N_i$ we denote the particle variations and $\alpha_{ij}$'s are called the FES parameters \cite{PhysRevLett.67.937.1991.Haldane}. 

The FES parameters must satisfy certain rules \cite{JPhysA.40.F1013.2007.Anghel,EPL.87.60009.2009.Anghel,PhysRevE.85.011144.2012.Liu,JStatMech.P04018.2013.Gundlach}, namely if we split an arbitrary species, $j$, into a number of sub-species, $j_0,j_1,\ldots$, then all the parameters $\alpha_{kl}$, with both, $k$ and $l$ different from $j$, remain unchanged, whereas the rest of the parameters must satisfy the relations:
\begin{subequations}\label{relalphasgen}
\begin{eqnarray}
\alpha_{ij} &=& \alpha_{ij_0}=\alpha_{ij_1}=\ldots,\ 
{\rm for\ any}\ i,\ i\ne j \label{alphaij} \\
\alpha_{ji} &=& \alpha_{j_0i}+\alpha_{j_1i}+\ldots,\ 
{\rm for\ any}\ i,\ i\ne j \label{alphaji} \\
\alpha_{jj} &=& \alpha_{j_0j_0}+\alpha_{j_1j_0}+\ldots = \alpha_{j_0j_1}+\alpha_{j_1j_1}+\ldots = \ldots
\label{alphajj}
\end{eqnarray}
\end{subequations}
These rules are satisfied by the ansatz \cite{JPhysA.40.F1013.2007.Anghel,EPL.90.10006.2010.Anghel},
\begin{subequations}\label{ansatz}
\begin{equation}
\alpha_{ij}=\alpha^{(e)}_{ij}+\alpha^{(s)}_{i}\delta_{ij} , \label{alpha_nd}
\end{equation}
where the parameters $\alpha^{(e)}_{ij}$, called the ``extensive'' parameters, are proportional to $G_i$,
\begin{equation}
\alpha^{(e)}_{ij}\equiv a_{ij}G_i. \label{def_alpha_e}
\end{equation}
\end{subequations}
The parameters $\alpha^{(s)}_{i}$ refer to only one species and do not depend on $G_i$.

The number of microscopic configurations compatible to a given distribution of particles on species, $\{N_i\}$, is 
\begin{equation}
  \WB(\{N_i\}) = \prod_i\frac{(G^{(-)}_i+N_i-1)!}{N_i!(G^{(-)}_i-1)!} \quad {\rm and} \quad \WF(\{N_i\}) = \prod_i \frac{G^{(+)}_i!}{N_i!(G^{(+)}_i-N_i)!} , \label{WBF_def}
\end{equation}
if the particles are bosons and fermions, respectively.

If for each species of particles, say species $i$, we associate an energy, $\epsilon_i$, and a chemical potential $\mu_i$, then the equilibrium particle distribution, $n^{(\pm)}_i\equiv N_i/G^{(\pm)}_i$, is obtained by maximizing the partition function, 
\begin{equation}
  \cZ^{(\pm)} \equiv \sum_{\{N_i\}}\left\{W^{(\pm)}(\{N_i\})\exp\left[\beta\sum_j(\epsilon_j-\mu_j)N_j\right]\right\} , \label{cZ_def}
\end{equation}
with respect to the distribution $\{N_i\}$, taking into account that the $G^{(\pm)}_j$'s vary with $\{N_i\}$ according to (\ref{alpha_def})--in Eq. (\ref{cZ_def}) $\beta=1/(\kB T)$ is the inverse temperature. 

The maximization of $\cZ^{(\pm)}$ with the conditions (\ref{alpha_def}) gives 
\begin{subequations}\label{Eqs_nis}
\begin{equation}
\beta(\mu_i-\epsilon_i)+\ln\frac{1\mp n^{(\pm)}_i}{n_i} = \mp\sum_{j} \alpha_{ji}\ln[1\mp n^{(\pm)}_{j}] .
 \label{inteq_for_n1_gen}
\end{equation}
where the all the upper signs are for fermions and all the lower signs are for bosons.

If one uses the ansatz (\ref{ansatz}) which is relevant for the systems analyzed below, Eq. (\ref{inteq_for_n1_gen}) becomes \cite{EPL.90.10006.2010.Anghel}
\begin{equation}
\beta(\mu_i-\epsilon_i)+\ln\frac{[1\mp n^{(\pm)}_i]^{1-\alpha^{(s)}_{i}}}
{n^{(\pm)}_i} = \mp\sum_{j} G_{j}a_{ji}\ln[1\mp n^{(\pm)}_{j}] .
 \label{inteq_for_n1}
\end{equation}
\end{subequations}

In FES, a system of fermions with a set of parameters, $\{\alpha_{ij}\}$, may be interpreted as a system of bosons with the parameters $\{\alpha_{ij}+\delta_{ij}\}$ and vice-versa, a system of bosons of parameters $\{\alpha_{ij}\}$ may be interpreted as a system of fermions with parameters $\{\alpha_{ij}-\delta_{ij}\}$. Therefore it is more natural to refer to Bose and Fermi \textit{formulations}, rather than to \textit{bosons} and \textit{fermions}. 

The most used formulation of FES is the one employed by Wu in Ref. \cite{PhysRevLett.73.922.1994.Wu}, which will be denoted here by ``W''. To see how this is related to the Bose formulation we define the number of states in the absence of particles in the system, $G^{\rm W}_i\equiv G^{(-)}_i+\sum\alpha_{ij}N_j$, and a new particle population, $n^{\rm W}_i\equiv N_i/G^{\rm W}_i$. Then the $n^{\rm W}_i$'s are determined in two steps. First one solves the system
\begin{subequations}\label{syst_Wu}
\begin{equation} 
  (1+w_{i}) \prod_{j}\left( \frac{w_{j}}{1+w_{j}} \right)^{\alpha_{ji}}
    = e^{(\epsilon_{i}-\mu)/\kB T}, \label{NLS}
\end{equation}
to determine the $w_i$'s, and then the $n_{i}^{\rm W}$'s are calculated from
\begin{eqnarray} 
  \sum_j(\delta_{ij}w_j + \alpha_{ij}G^{\rm W}_{j}/G^{\rm W}_{i})n_j = 1. \label{LS} 
\end{eqnarray} 
\end{subequations}
Comparing Eqs. (\ref{inteq_for_n1_gen}) and (\ref{NLS}) we observe that $w_i\equiv 1/n^{(-)}_i$.

Equations (\ref{Eqs_nis}) admit solutions of the Fermi liquid form \cite{PhysLettA.377.2922.2013.Anghel}
\begin{equation}
  n^{(\pm)}_i = \frac{1}{e^{\beta(\tilde\epsilon_i-\mu)}\pm 1} , \label{ansatz_ni}
\end{equation}
where $\tilde\epsilon_i$ are Landau type of quasiparticle energies that satisfy the general relations
\begin{eqnarray}
  \tilde\epsilon_k &=& \epsilon_k \mp k_BT \sum_i\alpha^{(\pm)}_{ik}\ln\left[1\mp n^{(\pm)}_i\right] . \nonumber \\
  &=& \epsilon_k \pm k_BT \sum_i\alpha^{(\pm)}_{ik}\ln\left[1\pm e^{-\beta(\tilde\epsilon_i-\mu)} \right]. \label{eq_tilmu1}
\end{eqnarray}

\section{Model and formalism} \label{model}


The particles are localized on random sites in a solid $s$-dimensional matrix. The positions of the sites are denoted by $\rr_I$, $I=1,2,\ldots,N_0$, where $N_0$ is the total number of sites. We assume that the wavefunctions of the particles do not overlap and the total energy of the system is 
\begin{equation}
  E = \sum_I \epsilon_{\rr_I} n_{\rr_I} + \frac{1}{2}\sum_{I,J}V_{\rr_I\rr_J}n_{\rr_I}n_{\rr_J}. \label{E_tot}
\end{equation}
where $\epsilon_{\rr_I}$ is the energy and $n_{\rr_I}$ is the occupation number of the site $I$; the total particle number is $N=\sum_{I}n_{\rr_I}$.

We shall work in the continuous limit, so we define the density of sites, $\sigma(\rr,\epsilon)\equiv\sum_I\delta^s(\rr-\rr_I)\delta(\epsilon-\epsilon_{\rr_I})$, and the particle density, $\rho(\rr,\epsilon)\equiv \sum_I \delta^s(\rr-\rr_I) \delta(\epsilon-\epsilon_{\rr_I}) n_{\rr_I}$.
The average particle population in an arbitrary $(s+1)$D volume, $\delta\Omega\times\delta\epsilon$, is 
$n(\rr,\epsilon) =[\int_{\delta\Omega}d^s\rr \int_{\delta\epsilon}d\epsilon\,\rho(\rr,\epsilon)] /[\int_{\delta\Omega}d^s\rr \int_{\delta\epsilon}d\epsilon\,\sigma(\rr,\epsilon)]$, where we assume that the volume is large enough, so that we have $\int_{\delta\Omega}d^s\rr\int_{\delta\epsilon}d\epsilon\,\sigma(\rr,\epsilon)\ge1$. In these notations the total energy of the system (\ref{E_tot}) becomes
\begin{eqnarray}
  E &=& \int_\Omega d^s\rr\int_\emin^\emax \epsilon \rho(\rr,\epsilon)\,d\epsilon \nonumber \\  
&& + \frac{1}{2}\int_\Omega d^s\rr\int_\Omega d^s\rr' \int_\emin^\emax d\epsilon \int_\emin^\emax d\epsilon' \rho(\rr,\epsilon)\rho(\rr',\epsilon')V_{\rr \rr'} \label{E_tot_int}
\end{eqnarray}
where $\Omega$ is total the volume of the system and $[\emin,\emax]$ is the interval in which $\epsilon$ takes values, with $\emin\ge 0$. 
We shall assume that $\emin=0$, $\emax\gg k_BT$ (where $k_B$ is the Boltzmann constant and $T$ is the temperature) and the interaction energy depends only on the distance between the sites, i.e.  $V_{\rr_I\rr_J}\equiv V(|\rr_I-\rr_J|)$. Because $\emax\gg k_BT$, we shall take $\emax=\infty$. 
For concreteness, we analyze only Fermi system, but the formalism can be easily extended to bosons.

To apply FES, we have to divide the system into species. We do this by coarse-graining the parameters space of the system, $\Omega\times\epsilon$, into the elementary volumes, $\delta\Omega_\xi\times\delta\epsilon_i$. By the lower case Greek letters, e.g. $\xi=0,1,\ldots$, we identify the spatial volume and by the lower case Latin letters, e.g. $i=0,1,\ldots$, we identify the energy intervals, $\delta\epsilon_i\equiv[\epsilon_i,\epsilon_{i+1}]$. We take $\epsilon_0=0$. 

We identify a species either directly, by $\delta\Omega_\xi\times\delta\epsilon_i$, or by the subscripts, 
$(\xi,i)$. We assume that each elementary volume, $\delta\Omega_\xi$, is centered at $\rr_\xi$ and 
contains a large enough number of sites and particles to justify the application of the statistical 
methods and, in particular the Stirling approximation for the logarithms of factorial numbers. In each of 
the volumes, say $\delta\Omega_\xi$, we have a distribution of sites, $\rr_I\in\delta\Omega_\xi$, of 
energies $\epsilon_{\rr_I}$. Under the assumption that the number of sites is large enough, the set of 
energies $\{\epsilon_{\rr_I}\}_{\rr_I\in\delta V_\xi}$ form a (quasi)continuous distribution along the 
$\epsilon$ axis, with a density $\sigma_\xi(\epsilon)\equiv\int_{\delta\Omega_\xi}\sigma(\rr,\epsilon)d^s\rr$. The number of states in the species 
$(\xi,i)$ is then $G_{\xi i}=\sigma_\xi(\epsilon_i)\delta\epsilon_i$ 
and the number of particles is $N_{\xi i}=\sigma_\xi(\epsilon_i)\delta\epsilon_i n(\rr_\xi,\epsilon_i)\equiv\delta \Omega_\xi\delta\epsilon_i\rho(\rr_\xi,\epsilon_i)$. 
We shall use the notations $\rho_\xi(\epsilon)\equiv\rho(\rr_\xi,\epsilon)$ 
for the particle density and  $\rho_{\xi}\equiv\int d\epsilon \rho(\rr_\xi,\epsilon)$
for the {\it volume} particle density.

We define the quasiparticle energies, $\tilde\epsilon_{\rr_I}$, in a similar way as in Ref. \cite{PhysLettA.372.5745.2008.Anghel,PhysLettA.376.892.2012.Anghel,PhysRevE.88.042150.2013.Anghel}, by
\begin{equation}
  \tilde\epsilon_{\rr_I} = \epsilon_{\rr_I} + \sum_{\tilde\epsilon_{\rr_J}<\tilde\epsilon_{\rr_I}} V(|\rr_I-\rr_J|)n_{\rr_J}. \label{def_qp_en1}
\end{equation}
Because of the identity $E=\sum_I \tilde\epsilon_{\rr_I}n_{\rr_I}$, the thermodynamics of the quasiparticle gas follows identically the thermodynamics of the original system. 
In the continuous limit, Eq. (\ref{def_qp_en1}) becomes
\begin{equation}
  \tilde\epsilon_{\rr_I} = \epsilon_{\rr_I} + \int_\Omega d^s\rr\int_0^{\epsilon_{\rr_I}}d\epsilon V(|\rr_I-\rr|)\sigma(\rr,\epsilon)n(\rr,\epsilon) . \label{def_qp_en1_int}
\end{equation}

We have a new parameters space, $\Omega\times\tilde\epsilon$, in which, by construction, $\temin=\emin=0$ and $\temax=\emax=\infty$. Using Eq. (\ref{def_qp_en1_int}) we obtain the new DOS,
\begin{equation}
  \tilde\sigma[\rr,\tilde\epsilon(\epsilon)] = \sigma(\rr,\epsilon)\left|\frac{d\tilde\epsilon}{d\epsilon}\right|^{-1} = \frac{\sigma(\rr,\epsilon)}{\left|1+\int_\Omega d^s\rr' V(|\rr-\rr'|)\sigma(\rr',\epsilon)n(\rr',\epsilon)\right|}. \label{tilde_sigma_def}
\end{equation}

\begin{figure}[t]
\begin{center}
\includegraphics[width=7.5cm]{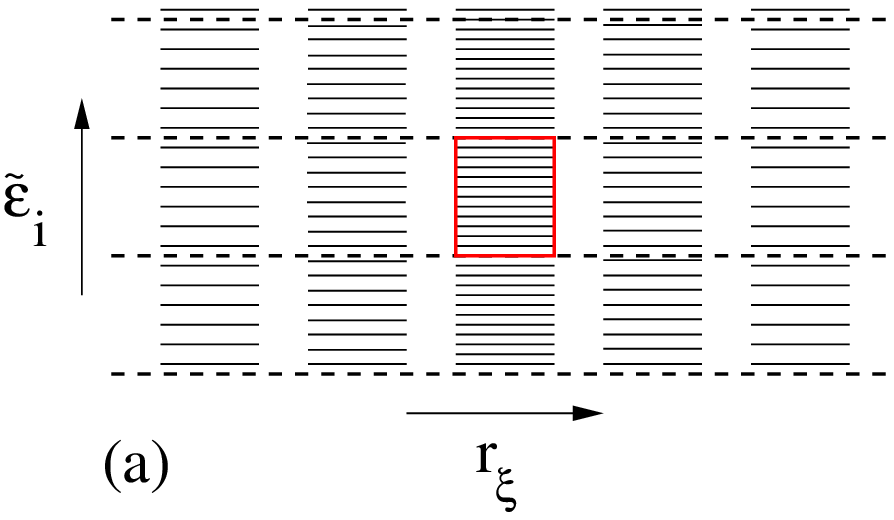}\\ 
\includegraphics[width=7.5cm]{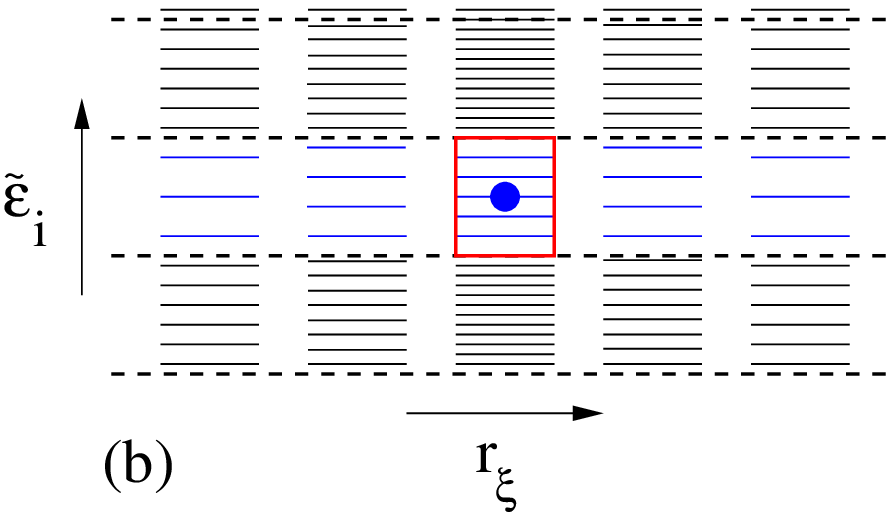}
\end{center}
\caption{
Partitioning of real space and quasiparticle energy axis. 
Upon inserting an extra particle in species $(\xi,i)$,
the numbers of available states in all species $(\eta,i)$ are changed.}
\label{species}
\end{figure}

Assuming that Eq. (\ref{def_qp_en1_int}) defines a one-to-one function $\tilde\epsilon(\epsilon)$ -- which may also be inverted to $\epsilon(\tilde\epsilon)$ -- we split the space $\Omega\times\tilde\epsilon$ into species, $\delta\Omega_\xi\times\delta\tilde\epsilon_i$, as we did with $\Omega\times\epsilon$, in such a way that $\tilde\epsilon_i\equiv\tilde\epsilon(\epsilon_i)$ for any $i$ and each species contains $G_{\xi i}$ states and $N_{\xi i}$ particles, as before. 
By the application of the procedure of Refs. \cite{PhysLettA.372.5745.2008.Anghel,PhysLettA.376.892.2012.Anghel} to this interaction potential and particle species we obtain the FES parameters
\cite{JPhysConfSer.410.012120.2013.Nemnes,PhysRevE.88.042150.2013.Anghel}
\begin{eqnarray}
  \alpha_{\xi i;\eta j} &=& \left[\delta_{ij}\sigma_\xi(\epsilon_i) + \theta(i-j) \delta\epsilon_i \frac{d\sigma_\xi(\epsilon_i)}{d\epsilon_i}\right] V(|\rr_\xi-\rr_\eta|) \nonumber \\ 
  &\equiv& \left[\delta_{ij}\delta\Omega_\xi\sigma(\rr_\xi,\epsilon_i) + \theta(i-j) \delta\epsilon_i \delta\Omega_\xi \frac{\partial\sigma(\rr_\xi,\epsilon_i)}{\partial\epsilon_i}\right] V(|\rr_\xi-\rr_\eta|) \label{alphas_def1}
\end{eqnarray}
where the first doublet, $(\xi i)$, specifies the species in which the number of states changes, whereas the second doublet, $(\eta j)$, specifies the species in which the number of particle changes; $\theta(k)$ is the step function, $\theta(k>0)=1$ and $\theta(k\le 0)=0$. The manifestation of the FES parameters given by Eq. (\ref{alphas_def1}) for a system with $\sigma(\rr,\epsilon)\equiv\sigma(\rr)$, i.e. independent of $\epsilon$, between species of the same quasiparticle energies is represented in Fig. \ref{species}.

We observe that the FES parameters (\ref{alphas_def1}) obey the rules (\ref{relalphasgen}) \cite{EPL.87.60009.2009.Anghel} and define a new ansatz, 
which is a generalization of (\ref{ansatz}) \cite{EPL.90.10006.2010.Anghel}.

The fact that the definition species, quasiparticle energies and FES parameters is self-consistent and feasible was checked in Ref. \cite{EPL.90.10006.2010.Anghel} where it was shown that the FES formalism is equivalent to the standard FLT for a general class of mean-field systems which includes also a non-constant external potential.

\subsection{Equilibrium thermodynamics}

Since we have fermions in the systems, we employ the Fermi formulation \cite{EPL.90.10006.2010.Anghel} to calculate the equilibrium thermodynamics. 
Plugging the $\alpha$ parameters (\ref{alphas_def1}) into the equations (\ref{Eqs_nis}) we get
\begin{eqnarray}
  0 &=& \beta(\mu-\tilde\epsilon_i) + \ln\frac{1-n^{(+)}(\rr_\xi,\tilde\epsilon_i)}{n^{(+)}(\rr_\xi,\tilde\epsilon_i)} + \sum_{\eta j} \alpha_{\eta j;\xi i} \ln[1-n^{(+)}(\rr_\eta,\tilde\epsilon_j)]. \label{Eq_d_equil_F}
\end{eqnarray}
In the continuous limit, Eq. (\ref{Eq_d_equil_F}) becomes
\begin{eqnarray}
  0 &=& \beta(\mu-\tilde\epsilon) + \ln\frac{1-n^{(+)}(\rr,\tilde\epsilon)}{n^{(+)}(\rr,\tilde\epsilon)} + \int_\Omega d^s\rr' V(|\rr'-\rr|) \sigma[\rr',\epsilon(\tilde\epsilon)]\ln[1-n^{(+)}(\rr',\tilde\epsilon)] \nonumber \\
  && + \int_\Omega d^s\rr' V(|\rr'-\rr|) \int_{\tilde\epsilon}^\infty d\tilde\epsilon'\left.\frac{\partial\sigma(\rr',\epsilon)}{\partial\epsilon}\right|_{\epsilon(\tilde\epsilon')} \ln[1-n^{(+)}(\rr',\tilde\epsilon')]. \label{Eq_equil_F}
\end{eqnarray}

In the bosonic formulation, as we mentioned above, $\alpha^{(-)}_{\xi i;\eta j} \equiv \delta_{\xi\eta}\delta_{ij}+\alpha_{\xi i;\eta j}$, and the dimension of the species is the number of available states, $G^{(-)}_{\xi i}\equiv G_{\xi i}-N_{\xi i}+1$ (we used the notation $G^{(-)}_{\xi i}$ to avoid confusion with the number of states in the species, $G_{\xi i}$). This changes also the definition of the population to $n^{(-)}(\rr_\xi,\epsilon_i)\equiv N_{\xi i}/G^{(-)}_{\xi i}\equiv n^{(+)}(\rr_\xi,\tilde\epsilon_i)/[1-n^{(+)}(\rr_\xi,\tilde\epsilon_i)]$. Plugging the new quantities into Eqs. (\ref{Eqs_nis}) we obtain the system of equations for $n^{(-)}(\rr_\xi,\epsilon_i)$:
\begin{eqnarray}
  0 &=& \beta(\mu-\tilde\epsilon_i) + \ln\frac{1+n^{(-)}(\rr_\xi,\tilde\epsilon_i)}{n^{(-)}(\rr_\xi,\tilde\epsilon_i)} - \sum_{\eta j}\alpha^{(-)}_{\eta j;\xi i}\ln[1+n^{(-)}(\rr_\eta,\tilde\epsilon_j)] \label{Eq_d_equil_B}
\end{eqnarray}
and in the continuous limit,
\begin{eqnarray}
  0 &=& \beta(\mu-\tilde\epsilon) + \ln\frac{1 +n^{(-)}(\rr,\tilde\epsilon)}{n^{(-)}(\rr,\tilde\epsilon)} - \int_\Omega d^s\rr' V(|\rr'-\rr|) \sigma[\rr',\epsilon(\tilde\epsilon)]\ln[1+n^{(-)}(\rr',\tilde\epsilon)] \nonumber \\ 
  && - \int_\Omega d^s\rr' V(|\rr'-\rr|) \int_{\tilde\epsilon}^\infty d\tilde\epsilon'\left.\frac{\partial\sigma(\rr',\epsilon)}{\partial\epsilon}\right|_{\epsilon(\tilde\epsilon')} \ln[1+n^{(-)}(\rr',\tilde\epsilon')]. \label{Eq_equil_B}
\end{eqnarray}

In Wu's formulation \cite{PhysRevLett.73.922.1994.Wu} the dimension of the species is $G^W_{\xi i}=G_{\xi i} + \sum_{\eta j}\alpha_{\xi i;\eta j}N_{\eta j} \equiv G^{(-)}_{\xi i} + \sum_{\eta j}\alpha^{(-)}_{\xi i;\eta j}N_{\eta j}$ and $n^W(\rr_\xi,\epsilon_i)\equiv N_{\xi i}/G^W_{\xi i}$. The equilibrium population is calculated from 
\begin{subequations} \label{Wu_disorder}
\begin{equation} 
  (1+w_{\xi i})\prod_{\eta, j} 
    \left( \frac{w_{\eta j}}{1+w_{\eta j}} \right)^{\alpha^{(-)}_{\eta j; \xi i}}
    = e^{(\epsilon_{\xi,i}-\mu)/kT}, \label{NLS_disorder}
\end{equation}
and 
\begin{equation} 
  \sum_{\eta,j}\left(\delta_{\xi\eta}\delta_{ij} w_{\eta j} + \beta_{\xi i, \eta j}\right) n^W_{\eta j} = 1 , \label{LS_disorder} 
\end{equation}
\end{subequations}
where $\beta_{\xi i, \eta j} = \alpha^{(-)}_{\xi i, \eta j} G^W_{\eta j} / G^W_{\xi i} \equiv\delta_{\xi\eta}\delta_{ij} + \alpha_{\xi i; \eta j} G^W_{\eta j} / G^W_{\xi i}$ and $w_{\xi i}\equiv \left(n^{(-)}_{\xi i}\right)^{-1}$.

Having the quasiparticle populations, we can calculate any thermodynamical quantity. The internal energy of the system is 
\begin{equation}
  U(T,\mu) = \sum_{I} n_{\rr_I}\tilde\epsilon_{\rr_I} = \sum_{\xi i} n^{P}(\rr_\xi,\tilde\epsilon_i) G^{P}_{\xi i} \tilde\epsilon_i \label{U_def_qp}
\end{equation}
or, in the continuous limit, 
\begin{eqnarray}
 U(T,\mu) &=& \int_\Omega d^s\rr\int_0^\infty d\tilde\epsilon\,\tilde\sigma^{P}(\rr,\tilde\epsilon)n^{P}(\rr,\tilde\epsilon)\tilde\epsilon \equiv \int_\Omega d^s\rr\int_0^\infty d\tilde\epsilon\,\tilde\rho(\rr,\tilde\epsilon) \tilde\epsilon, \label{U_def_gen}
\end{eqnarray}
where $P=(+),\ (-)$ or $W$ and $\tilde\sigma^{(+)}(\rr,\tilde\epsilon)$ is given by Eq. (\ref{tilde_sigma_def}). 
The other two densities of states are $\tilde\sigma^{(-)}(\rr,\tilde\epsilon)\equiv\tilde\sigma^{(+)}(\rr,\tilde\epsilon)-\tilde\rho(\rr,\tilde\epsilon)$ and $\tilde\sigma^\rmW(\rr,\tilde\epsilon)$ and we have the relation $\tilde\rho(\rr,\tilde\epsilon)\equiv\tilde\sigma^{(+)}(\rr,\tilde\epsilon)n^{(+)}(\rr,\tilde\epsilon) \equiv\tilde\sigma^{(-)}(\rr,\tilde\epsilon)n^{(-)}(\rr,\tilde\epsilon) \equiv\tilde\sigma^W(\rr,\tilde\epsilon) n^W(\rr,\tilde\epsilon)$.
We shall use the notation $\tilde\rho_\xi(\tilde\epsilon)\equiv\tilde\rho(\rr_\xi,\tilde\epsilon)$
for the quasi-particle density in $\Omega_\xi$, at quasiparticle energy $\tilde\epsilon$, and $\tilde\rho_\xi\equiv\int d\tilde\epsilon\tilde\rho(\rr,\tilde\epsilon)$ for the {\it volume} quasi-particle density. One should note that $\tilde\rho_\xi\equiv\rho_\xi$ by definition. 

In order to define the DOS $\tilde\sigma^W(\rr,\tilde\epsilon)$ we have to define the densities of the FES parameters \cite{JPhysA.40.F1013.2007.Anghel,EPL.90.10006.2010.Anghel}, $a_{\rr\tilde\epsilon;\rr'\tilde\epsilon'}$, by
\begin{subequations} \label{a_defs}
\begin{eqnarray}
  \alpha_{\xi i;\eta j} &=& a_{\rr_\xi \tilde\epsilon_i;\rr_\eta \tilde\epsilon_j} \delta\epsilon_i \delta\Omega_\xi \label{a_def1} 
\end{eqnarray}
where 
\begin{eqnarray}
  a_{\rr_\xi \tilde\epsilon_i;\rr_\eta \tilde\epsilon_j} &=& \left[\delta(\tilde\epsilon_i-\tilde\epsilon_j) \sigma(\rr_\xi,\epsilon_i) + \theta(\tilde\epsilon_i-\tilde\epsilon_j) \frac{\partial\sigma(\rr_\xi,\epsilon_i)}{\partial\epsilon_i}\right] V(|\rr_\xi-\rr_\eta|) . \label{a_def2} 
\end{eqnarray}
\end{subequations}
Using Eqs. (\ref{a_defs}) we write $\tilde\sigma^W(\rr,\tilde\epsilon)=\tilde\sigma^{(+)}(\rr,\tilde\epsilon)+\int_\Omega d^s\rr'\int_0^\infty d\tilde\epsilon'\,a_{\rr\tilde\epsilon;\rr' \tilde\epsilon'} \tilde\rho(\rr',\tilde\epsilon')$.

Similarly to Eq. (\ref{U_def_gen}), the total particle number is
\begin{eqnarray}
 N(T,\mu) &=& \int_\Omega d^s\rr\int_0^\infty d\tilde\epsilon\,\tilde\sigma^{P}(\rr,\tilde\epsilon)n^{P}(\rr,\tilde\epsilon) \equiv \int_\Omega d^s\rr\int_0^\infty d\tilde\epsilon\,\tilde\rho(\rr,\tilde\epsilon). \label{N_def_gen}
\end{eqnarray}

The heat capacity and the entropy of the system are
\begin{eqnarray}
  C_V &=& \left(\frac{\partial U}{\partial T}\right)_N = \frac{\partial U(T,\mu)}{\partial T} - \frac{\partial U(T,\mu)}{\partial\mu}\frac{\partial N(T,\mu)}{\partial T} \left(\frac{\partial N(T,\mu)}{\partial\mu}\right)^{-1} \label{CV}
\end{eqnarray}
and 
\begin{equation}
S = k_B \ln(W^{(\pm)}),
\end{equation}
respectively, and they satisfy the equation
\begin{equation}
\label{CvS}
C_V = T\left(\frac{\partial S}{\partial T}\right)_N.
\end{equation}

\subsection{Homogeneous system}
\label{homsys}

If $\sigma(\rr,\epsilon)\equiv\sigma (\epsilon)$ is independent of $\rr$ and if we impose periodic boundary conditions or ignore the effect of the surfaces (deep inside the solid), then both, $n(\rr,\epsilon)$ and $\tilde\sigma(\rr,\tilde\epsilon)$, are independent of $\rr$ and in the Eqs. (\ref{Eq_equil_F}) and (\ref{Eq_equil_B}) we can perform the integrals over $\Omega$ to obtain an 
equation only in $\tilde\epsilon$:
\begin{eqnarray}
  0 &=& \beta(\mu-\tilde\epsilon) + \ln\frac{1-n^{(+)}(\tilde\epsilon)}{n^{(+)}(\tilde\epsilon)} + I_V \left\{ \sigma[\epsilon(\tilde\epsilon)]\ln[1-n^{(+)}(\tilde\epsilon)] \right. \nonumber \\
  && + \left. \int_{\tilde\epsilon}^\infty d\tilde\epsilon'\left.\frac{\partial\sigma(\epsilon)}{\partial\epsilon}\right|_{\epsilon(\tilde\epsilon')} \ln[1-n^{(+)}(\tilde\epsilon')] \right\} , \label{Eq_equil_F_hom}
\end{eqnarray}
for fermions or
\begin{eqnarray}
  0 &=& \beta(\mu-\tilde\epsilon) + \ln\frac{1 +n^{(-)}(\tilde\epsilon)}{n^{(-)}(\tilde\epsilon)} - I_V \left\{ \sigma[\epsilon(\tilde\epsilon)]\ln[1+n^{(-)}(\tilde\epsilon)] \right. \nonumber \\ 
  && \left. - \int_{\tilde\epsilon}^\infty d\tilde\epsilon'\left.\frac{\partial\sigma(\epsilon)}{\partial\epsilon}\right|_{\epsilon(\tilde\epsilon')} \ln[1+n^{(-)}(\tilde\epsilon')]\right\} , \label{Eq_equil_B_hom}
\end{eqnarray}
for bosons, where
\begin{equation}
  I_V = \int_\Omega d^s\rr' V(|\rr'-\rr|) . \label{int_V}
\end{equation}

Moreover, if $\sigma(\rr,\epsilon)\equiv\sigma$, i.e. it does not depend on energy, then one may define an effective FES parameter,
$\alpha_{\rm eff}(\tilde\epsilon) = \sigma I_V$, 
and Eqs. (\ref{Eq_equil_F_hom}) and (\ref{Eq_equil_B_hom}) simplify to
\begin{equation}
  \beta(\tilde\epsilon-\mu) = \ln\frac{[1-n^{(+)}(\rr,\tilde\epsilon)]^{1+\alpha_{\rm eff}}}{n^{(+)}(\rr,\tilde\epsilon)} \label{Eqs_equil_F_hom2}
\end{equation}
and
\begin{equation}
  \beta(\tilde\epsilon-\mu) = \ln\frac{[1 +n^{(-)}(\rr,\tilde\epsilon)]^{1-\alpha_{\rm eff}}}{n^{(-)}(\rr,\tilde\epsilon)} , \label{Eqs_equil_B_hom2}
\end{equation}
respectively.

\section{Applications}\label{applications}

\subsection{Two sub-volumes system}
\label{23sp}

First let's consider the simple example of a system formed of two sub-volumes, $\delta\Omega_0$ and $\delta\Omega_1$.
%
%
Such a test-case has the advantage of being transparent enough for a detailed discussion while
still providing significant insights into the thermodynamic of
more general interacting inhomogeneous systems. Moreover, for properly chosen parameters the populations may be calculated analytically.

We define the energy independent densities of states for the non-interacting particles in the two sub-volumes, $\sigma_0$ and $\sigma_1$, and if we take the interaction energies between the particles in the same volume to be $V_{00}=V_{11}$ and in different volumes $V_{01}=V_{10}$, then the FES parameters are $\alpha_{\xi i,\eta j}= \sigma_\xi V_{\xi \eta}\delta_{i j}$, where $\xi,\eta=0,1$ and $i,j$ denote the energy intervals, as explained above.
The total number of particles is defined as $N= 2 E_{\rm F} \sigma\equiv E_{\rm F}(\sigma_0+\sigma_1)$, where $\sigma$ is the average DOS and $E_{\rm F}$ is the Fermi energy for the non-interacting particle system. In this example we take $2\sigma_0=\sigma_1=1$.

\begin{figure}[t]
\begin{center}
\includegraphics[width=11.5cm]{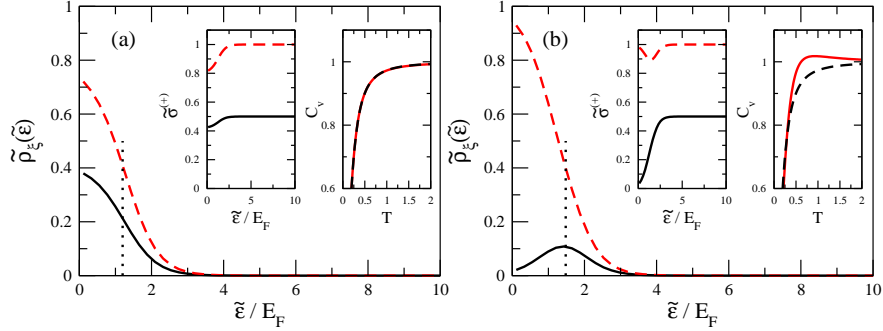}
\end{center}
\caption{(Color online) Main plots: particle density vs. quasiparticle energy 
for the system consisting of two subsystems of volumes $\delta\Omega_0$ and $\delta\Omega_1$, which can exchange particles between them. The black curves correspond to sub-system 0 ($\xi=0$) and the dashed, red curves correspond to the sub-system 1 ($\xi=1$). Panels 
(a) and (b) contain the results for the type 1 ($V_{00}=V_{11}=0.2>V_{01}=V_{10}=0.1$) and type 2 ($V_{00}=V_{11}=0<V_{01}=1.0$) systems, respectively.
In both panels, the left insets contain the fermionic DOS 
as functions of quasiparticle energy, for $\xi=0$ (black/solid) and $\xi=1$ (red/dashed), whereas
the right insets contain the temperature dependence of the heat capacity of the system
(red/solid) and of a non-interacting system of fermions of the same, constant DOS
(black/dashed).
The main plots and the left panels correspond to a temperature $\kB T=0.4 E_{\rm F}$, where $E_{\rm F}$ is the Fermi energy.
The chemical potentials are marked by vertical dotted lines.
}
\label{2_3_sp}
\end{figure}

We investigate two types of systems: type 1 (figure \ref{2_3_sp} a), with $0<V_{10}=V_{01}<V_{00}=V_{11}$ (since the particles in the same volume, being closer together, interact stronger than the particles in different volumes), and type 2 (figure \ref{2_3_sp} b), with the particles in the same volume behaving like ideal fermions, but interact repulsively with the particles in the other volume: $0=V_{11}=V_{00}<V_{01}=V_{10}$. 


In figure \ref{2_3_sp} (a) we observe that the particle densities in both volumes decrease with the quasiparticle energy, as one would expect, whereas, due to the repulsive interaction between the atoms, the fermionic densities of states increase monotonically. At high energies the fermionic densities of states converge to the non-interacting densities, $\sigma_0$ and $\sigma_1$. These curves are calculated for $\kB T=0.4 E_{\rm F}$, but they are typical for this system. In the right inset of figure \ref{2_3_sp} (a) we plot the heat capacity, which is not much different from the heat capacity of an ideal Fermi system.


The second example (figure \ref{2_3_sp} b) is equivalent to two subsystems of ideal fermions, with mutual interaction and which can exchange particles between them. This choice of parameters lead to $\alpha_{00}=\alpha_{11}=0$ and $2\alpha_{01}=\alpha_{10}$. Choosing $\alpha_{10}=1$ and denoting the fugacity by $g(\tilde\epsilon)=\exp[\beta(\tilde\epsilon-\mu)]$, we obtain from the system (\ref{Eq_d_equil_F}) a third order equation for $n^{(+)}_1(\tilde\epsilon)\equiv n^{(+)}_1(g)$,
\begin{subequations}\label{eq3drg}
\begin{equation}
  g \left[n^{(+)}_{1}(g)\right]^3+(1-g-g^2)\left[n^{(+)}_{1}(g)\right]^2 -2n^{(+)}_{1}(g)+1=0, \label{eq3drg_n1}
\end{equation}
and an equation for $n^{(+)}_0(\epsilon)$,
\begin{equation}
  n^{(+)}_{0}(g)=1-\frac{g^2\left[n^{(+)}_{1}(g)\right]^2}{\left[1-n^{(+)}_{1}(g)\right]^2} 
\label{eq3drg_n0}
\end{equation}
\end{subequations}
The third degree equation from (\ref{eq3drg_n1}) has three real,
distinct solutions for $g>0$. However, by imposing the conditions
$n^{(+)}_{\xi}(\tilde\epsilon)\ge0$ and $\lim_{\tilde\epsilon\rightarrow\infty}n^{(+)}_{\xi}(\tilde\epsilon)=0$, we remain with only one solution, which is represented in Fig.\ \ref{2_3_sp}(b). 

In contrast to the previous system, a maximum occurs 
in $\tilde\rho_0(\tilde\epsilon)$ at $\tilde\epsilon=\mu$,
indicating a population inversion with respect to the quasiparticle energy in the volume with lower DOS. This is due to the mutual exclusion statistics (\ref{alpha_def}), which reduces the density of states significantly at low energies, especially in the species 0, where the density of states is lower.

At larger quasiparticle energies, $\tilde\epsilon>\mu$, the densities of particles are lower and therefore the statistical interaction effects are also diminished.
It is worth mentioning that for the total particle density,
$\tilde\rho(\tilde\epsilon)=\tilde\rho_0(\tilde\epsilon)+\tilde\rho_1(\tilde\epsilon)$, 
no population inversion occurs, i.e. $\tilde\rho(\tilde\epsilon)$ is a monotonically decreasing function of the quasiparticle energy.
The maximum observed in $\tilde\rho_0(\tilde\epsilon)$ induces a
minimum in $\tilde\sigma_1^{(+)}(\tilde\epsilon)$, whereas $\tilde\sigma_0^{(+)}$ increases monotonically with $\tilde\epsilon$. 

Systems of particles with the same constant DOS and direct FES parameters, $\alpha_{ij}=\alpha\delta_{ij}$, have the same heat capacity, which is independent of $\alpha$ and which is monotonically increasing with $T$ \cite{PhysRevE.75.61120.2007.Potter,ProcCambrPhilos42.272.1946.Auluc,PhysRev.135.A1515.1964.May,PhysRevE.55.1518.1997.Lee,JPA35.7255.2002.Anghel}. In our case the existence of mutual FES parameters, $\alpha_{ij}\ne 0$ for $i\ne j$, not only changes the heat capacity, but also leads to the appearance of a maximum in $C_{\rm v}(T)$. For $T\to\infty$, $C_{\rm v}(T)$ converges to 1, which is the Boltzmann limit, for any FES parameters.


\subsection{Screened Coulomb interactions. Homogeneous systems.} \label{Ci_HS}

We further assume a two-particle screened potential of the general form
\begin{equation}
V(r;\gamma,\lambda) = {\kappa} \frac{\exp(-r/\lambda)}{r^\gamma}, \label{V_model}
\end{equation}
where $r=|\bf{r}-\bf{r'}|$. In particular, if $\gamma=1$, we have the usual screened Coulomb and Yukawa type potentials. In the absence of screening ($\lambda\rightarrow\infty$), such systems exhibit standard thermostatistical behavior if $d/\gamma<1$
\cite{ChaosSolitonsFractals.13.371.2002.Tsallis} and the interactions are classified as short ranged. If $d/\gamma \ge 1$ the systems obey {\it non-extensive} thermodynamics, i.e. quantities like total energy are not extensive due to the long range interactions. 
However if the screening is present, the interactions become short-ranged and
the usual thermodynamics applies. 

To remove the singularity at the origin that appear in the integrals over $V({\rm r})$ in Eqs. (\ref{Eq_equil_F}), (\ref{Eq_equil_B}) and (\ref{int_V}), we introduce a cut-off at radius $R_0$, below which the potential remains constant -- $V(r) = V(R_0) \equiv V_0$ for $r<R_0$. With this assumption, for a homogeneous system, 
\begin{equation}
\label{IV}
  I_V = \frac{2\pi^\frac{s}{2}}{\Gamma\left(\frac{s}{2}\right)} \int_0^{\infty} dr \; r^{s-1}V(r) = I_{V1} + I_{V2},
\end{equation}
where
\begin{eqnarray}
\label{alpha12}
\nonumber
  I_{V1} &\equiv& \frac{2\pi^\frac{s}{2}}{\Gamma\left(\frac{s}{2}\right)} \int_0^{R_0} dr \; r^{s-1}V(r) = \frac{\pi^\frac{s}{2}}{\Gamma\left(\frac{s}{2}+1\right)} R_0^s V_0 \\
  I_{V2} &\equiv& I_V-I_{V1} = \frac{2\pi^\frac{s}{2}}{\Gamma\left(\frac{s}{2}\right)} \kappa \sigma \lambda^{s-\gamma} \Gamma\left(s-\gamma,\frac{R_0}{\lambda}\right). 
\end{eqnarray}
Here $\Gamma(s,x)=\int_x^\infty t^{s-1} e^{-t} dt$ is the upper incomplete gamma function.

For a screened Coulomb-type interaction in a $s$-dimensional system ($s=1,2,3$), 
the term $I_{V2}$ can be
expressed as:
\begin{eqnarray}
\label{alpha2}
\nonumber
I_{V2}^{1D} &=& 2\kappa\sigma E_1(R_0/\lambda)\\
\nonumber
I_{V2}^{2D} &=& 2\pi\kappa\sigma \;
                    \lambda\exp\left(-\frac{R_0}{\lambda}\right)\\
I_{V2}^{3D} &=& 4\pi\kappa\sigma \; 
                    \lambda(R_0+\lambda)\exp\left(-\frac{R_0}{\lambda}\right),
\end{eqnarray}
where $E_1(z)$ is the exponential integral,
\begin{equation}
\label{E1}
  E_1(z) = \int_z^\infty \frac{e^{-t}}{t} dt
         = \int_1^\infty \frac{e^{-zt}}{t} dt. 
\end{equation}
Using the integrals (\ref{alpha2}) calculated analitically one can apply the formalism presented in Subsection \ref{homsys} and therefore the complete
thermodynamical behavior of the homogeneous system may be calculated.

\subsection{Accumulation and depletion effects in one-dimensional systems}
\label{app}

Using the FES formalism we next describe the accumulation and depletion effects 
for one dimensional systems of particles with screened Coulomb interactions
and non-uniform DOS.
Periodic boundary conditions are imposed using the minimum image 
convention, i.e. the interaction drops at a distance equal to half the
length of the repetitive unit \cite{PhilMagB.81.1117.2001.Vojta}.

For simplicity we shall assume in the following that $\sigma(\rr,\epsilon)$ [and $\sigma_\xi(\epsilon)$] are independent of $\epsilon$, so we shall simplify the notation of the DOS to $\sigma(\rr)$ (and $\sigma_\xi$). In this case the FES parameters (\ref{alphas_def1}) can be written as
\begin{equation}
  \alpha_{\xi i;\eta j} = \delta_{ij}\delta\Omega_\xi\sigma(\rr_\xi,\epsilon_i) V(|\rr_\xi-\rr_\eta|) \equiv \delta_{ij}\delta\Omega_\xi\sigma_\xi(\epsilon_i) V_{\xi\eta}, \label{alphas_def1_diag}
\end{equation}
which are diagonal in the indices $i$ and $j$. 

For our 1D system we apply a spatial partitioning such that $\delta\Omega_\xi \equiv R_\xi = 1$ and we take the cut-off distance, $R_0$, to be half of the species dimension, i.e. $R_0=R_\xi/2$. Furthermore, we set $V(R_\xi)=1$, which implies $V(R_0)=2$. Under these assumptions we rewrite the FES parameters in Eq. 
(\ref{alphas_def1_diag}) as:
\begin{eqnarray}
\nonumber
\alpha_{\xi i;\xi i} &=& \sigma_\xi(\epsilon_i) V_0 R_\xi \\
\nonumber
\alpha_{\xi i;\eta j} &=& \delta_{ij}\sigma_\xi(\epsilon_i) 
                  (V_0/2) R_\xi^2/|\rr_\xi-\rr_\eta| \;\;\; \mbox{for} \;\;\; i \neq j.
\end{eqnarray}

The quasiparticle densities $\tilde\rho_{\xi}(\tilde\epsilon)$, $\tilde\rho_\xi$ can be found
in any of the descriptions aforementioned (fermionic, bosonic and $W$) by solving the corresponding nonlinear system (\ref{Eq_equil_F}), (\ref{Eq_equil_B}) or (\ref{Wu_disorder}). The solution is found iteratively using gradient descent method -- as implemented by GSL \cite{gsl} -- starting from an initial guess solution. 
A key point is related to the solver initialization. A feasable solution may
be represented by the equilibrium population of the non-interacting system.
However, a more rapid convergence is obtained once the solver is initialized
using the solution obtained using Eqs. (\ref{IV}-\ref{E1}) for 
a homogeneous system with the average DOS and the same interaction, 
where simpler calculations can be performed, as indicated in Section \ref{Ci_HS}.
Since the thermodynamical quantities of interest should be obtained for
a given particle number $N$ rather than for a given $\mu$, 
an extra loop is introduced so that the corresponding chemical potential 
is found in agreement with the normalization relation (\ref{N_def_gen}).

\begin{figure}[t]
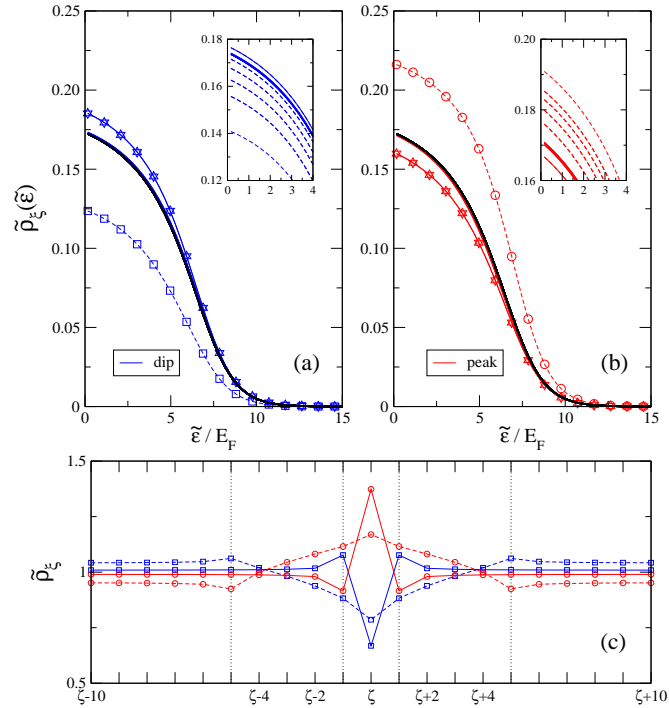

\begin{center}
\includegraphics[width=8.5cm]{figure3a_3b.eps} \\
\hspace*{0.33cm}\includegraphics[width=8.4cm]{figure3c.eps}
\end{center}
\caption{(Color online) Quasiparticle density, $\tilde\rho_{\xi}(\tilde\epsilon)$, for systems with 
a {\it dip} ($\sigma_\zeta=0.5$) (a) and a {\it peak} ($\sigma_\zeta=2.0$) (b) 
in the otherwise constant DOS
($\sigma_\xi=1.0$, for $\xi \neq \zeta$). 
The symbols represent the particle densities which correspond to the species 
$\zeta$ [squares (a) and circles (b)] and its first nearest neighbors (up/down triangles).
The solid black lines correspond to a uniform systems with $\sigma = 1.0$. 
The insets contain similar data for systems with a finite width dip/peak.
The dashed lines represent particle densities for sub-volumes
$[\zeta-\tau,\zeta+\tau]$, with $\tau \le 4$.
The volume density of (quasi)particles $\tilde\rho_{\xi}$ 
in each species are represented in (c). 
Pairs of vertical 
dotted lines mark the extension of the considered dips and peaks.}
\label{rho_dip_peak}
\end{figure}

We consider a system with 
repulsive screened Coulomb interactions, parameterized by 
$\gamma=1$ and $\lambda=3R_\xi$. 
For the partition of the physical space and the coarse graining on the 
energy axis we take $N_R=20$ and $N_E=50$, respectively.
The system contains a number of $N = N_R E_F \sigma$ particles,
where $\sigma$ is the average DOS.

\begin{figure}[t]
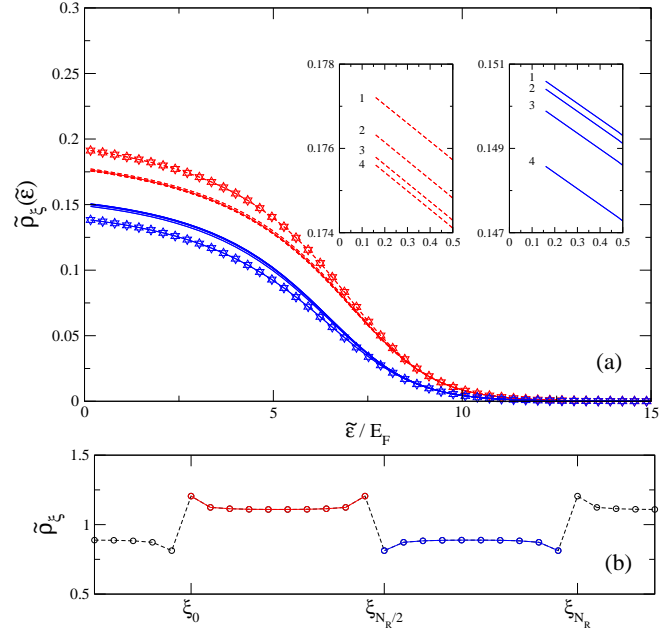

\begin{center}
\includegraphics[width=8.5cm]{figure4a.eps}\\
\hspace*{0.25cm}\includegraphics[width=8.25cm]{figure4b.eps}
\end{center}
\caption{(Color online) (a) Quasiparticle density for a system with a step-like DOS,
$\sigma_\xi=1$, for $\xi<N_R/2$ (upper curves, in red) and
$\sigma_\xi=0.5$, for $\xi \ge N_R/2$ (lower curves, in blue).
The symbols (up/down triangles) correspond to the four sub-volumes located 
at the interfaces. In the insets (left, for $\xi < N_R/2$, and right, 
for $\xi \ge N_R/2$) are shown details of 
$\tilde\rho_{\xi}(\tilde\epsilon)$
for the remaining sub-volumes. The numbers indicate the minimum distance 
from either interface.
(b) Spatial (quasi)particle density $\tilde\rho_\xi$ of the considered system.}
\label{rho_interface}
\end{figure}

Two particularly transparent examples of an inhomogeneous system 
are obtained by introducing a dip or a 
peak in the density of states of a homogeneous system. More concretely, 
we choose one sub-volume, $\zeta$, for which the density of states is half or
twice the reference (constant) value of the other sub-volumes --
$\sigma_\xi = 1.0$ for any $\xi \neq \zeta$ and 
$\sigma_\zeta = 0.5$ (``dip'' case) or $2.0$ (``peak'' case).
For symmetry reasons we have considered $N_R+1$ sub-volumes.
Figure \ref{rho_dip_peak} depicts the particle densities for
the two types of systems considered at $T=E_{\rm F}/\kB$. Unlike a homogeneous system, 
where there is a unique function $n_{\xi i} \equiv n_i$ for all sub-volumes,
corresponding to a (single) mean field FES parameter $\alpha_{\rm eff}$,
we now have different populations with a spatial distribution.
As one can see from Fig.~\ref{rho_dip_peak} 
the systems obey a mirrored symmetry:
the density of particles in
sub-volume $\zeta$ drops in the ``dip'' case and is enhanced in ``peak'' case, as compared with a homogeneous system, $\sigma_\xi=1.0$ for any $\xi$.
The particle densities in nearest neighbors sub-volumes of $\zeta$, 
i.e. $\zeta-1$ and $\zeta+1$,
exhibit deviations from the mean field values and opposite to the values in
sub-volume $\zeta$. This can be explained as follows.

At equilibrium, in the case of a dip in the density of states, 
the repulsive interaction between particles drives accumulations of particles
towards the edges of the region with constant $\sigma=1.0$,
as the number of particle in sub-volume $\zeta$ is lower due to a smaller
number of available states. 
The opposite, namely a depletion of particles, 
is found in the case of a peak in the DOS, where the larger 
number of particles in sub-volume $\zeta$ repel the particles in the 
adjacent regions. 
Starting with the $\zeta-2$ and $\zeta+2$ sub-volumes (two sub-volumes away), the populations
already get very close to the values corresponding to the homogeneous system.

However, for dips or peaks of finite widths,
a broader distribution of particle densities can be observed in the insets 
of Fig.~\ref{rho_dip_peak}. Here we took the linear dependences
$\sigma_{\zeta+\tau} = \sigma_{\zeta-\tau} = 0.5 + 0.1\tau$ (dip)
and $\sigma_{\zeta+\tau} = \sigma_{\zeta-\tau} = 2.0 - 0.2\tau$ (peak),
with $\tau=1,2,3,4$.
Note that due to the symmetry, except the $\zeta$ sub-volume, we have 
the pairs $\rho_{\zeta-\tau, i} = \rho_{\zeta+\tau, i}$, for any $\tau>0$. 

We further analyze the particle distribution in a periodic 
one-dimensional system with interfaces
between high ($\sigma_\xi=1.0$, for $\xi<N_R/2$) 
and low ($\sigma_\xi=0.5$, for $\xi \ge N_R/2$) DOS regions. 
The obtained results are represented in Fig.\ \ref{rho_interface} for 
a temperature $T=E_{\rm F}/\kB$.
From the main plot one can see the particle densities are divided in two
groups corresponding to the two values of $\sigma$. 
Like in the previous case, significant deviations occur 
at the interfaces. 
The two sub-volumes adjacent to the interfaces indicate accumulations and depletions
of particles, determined by the high and low DOS values, respectively.  
The particle densities in the other sub-volumes are depicted in detail in 
the two insets, indicating a convergence towards the values corresponding to 
the mid-points of the two homogeneous regions.
The exact extent of the deviations in particle distributions compared to the homogeneous system depends also on the type of interacting potential and it can be accurately described by the FES formalism.

\begin{figure}[t]
\begin{center}
\includegraphics[width=8.5cm]{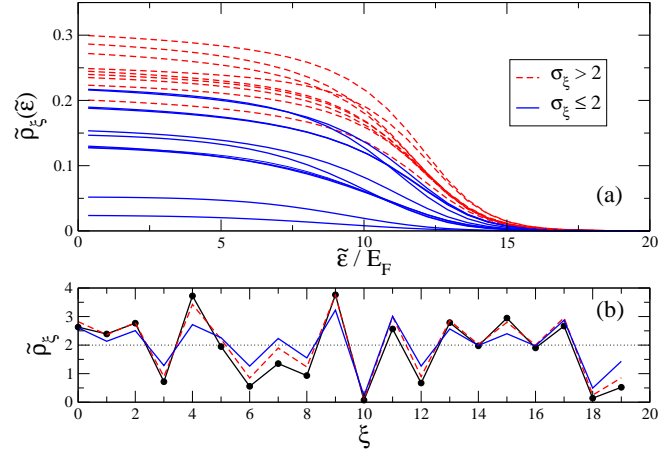}
\end{center}
\caption{(Color online) (a) Quasiparticle density for a disordered system 
($\sigma_0=2$, $\Delta\sigma = 4$).
The species with high (blue/solid) and 
low (red/dashed) DOS are indicated.
(b) Density of states $\sigma_\xi$ (black/solid with dots) 
and particle numbers in 
each sub-volume for a high ($T_h=6.0$) (red/dashed) and 
a low ($T_l=0.5$) (blue/solid) temperature.
}
\label{rho_random}
\end{figure}

\subsection{Quasiparticle segregation and thermodynamical consequences in 1D disordered systems.}
\label{qpartseg}

We analyze in the following the properties of 
a system with disorder, which we introduce by 
randomly distributed values for the local DOS $\sigma_\xi$
at each site. We consider a step distribution centered around
the average value $\sigma_0 = 2$ and several degrees of disorder determined by the width of the distribution, $\Delta\sigma$.
Figure \ref{rho_random}(a) shows a typical particle distribution
obtained at a temperature $T=E_{\rm F}/\kB$ and maximum disorder, 
$\Delta\sigma=4$. 

Like in the two previously analyzed examples one can 
observe the particle densities are larger at sites with larger DOS,
although now we find the disorder specific distributions $\rho_{\xi i}$.  
In the lower plot is represented the local DOS together with the particle
distributions at each site $\xi$, for two temperatures $T_l=0.5E_{\rm F}/\kB$ and $T_h=6E_{\rm F}/\kB$.
One can see that for the higher temperature $T_h$ the particle distribution 
follows closely the $\sigma_\xi$ distribution, while for the the lower
temperature $T_l$ the particles are more evenly distributed. This is because
in the high temperature limit the interactions become less and less
important, the particle distributions approach the Maxwell-Boltzmann 
distribution and all available states become equally probable. 
Consequently, the spatial particle density $\rho_\xi$ in each sub-volume 
becomes proportional to the density of states $\sigma_\xi$.
By contrast, in the lower temperature limit, the repulsive interactions
tend to level the particle distribution. 
At low and intermediate temperatures, 
there is no clear relation between the particle distribution and the local DOS.

\begin{figure}[t]
\begin{center}
\includegraphics[width=8.5cm]{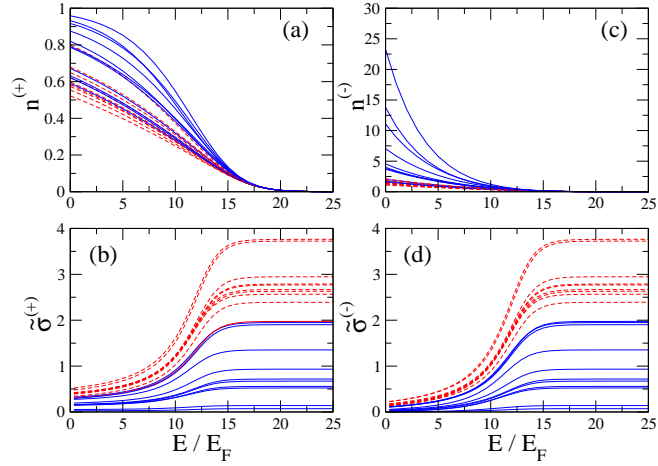}
\end{center}
\caption{(Color online)
Populations and quasi-particle DOS
(red/dashed for $\sigma_\xi>2$, blue/dashed for $\sigma_\xi\le2$)
in the fermionic (a-b) and bosonic (c-d) descriptions.}
\label{nFB}
\end{figure}

\begin{figure}[t]
\begin{center}
\includegraphics[width=8.5cm]{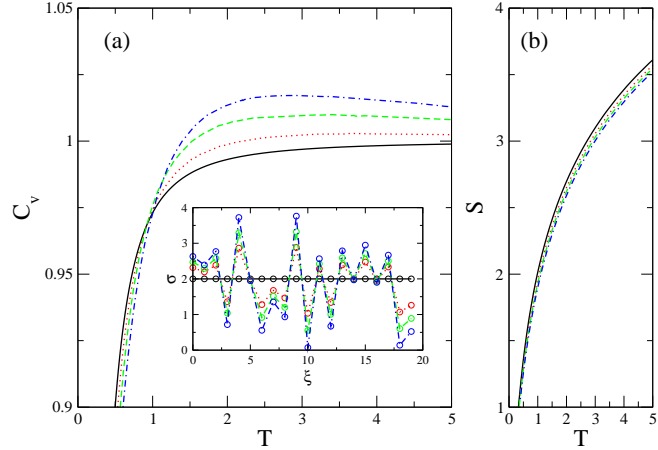}
\end{center}
\caption{
(Color online)
Heat capacity (a) and entropy (b) for different degrees of disorder:
$\Delta\sigma=0$ (black/solid), $0.5$ (red/dotted),
$0.75$ (green/dashed) and $1$ (blue/dashed-dotted), 
as indicated in the inset.}
\label{Cv_T_20_50}
\end{figure}

The results presented so far for the particle density distributions may be obtained in
any of the three descriptions mentioned in the section \ref{model}, 
i.e. the {\it fermionic}, {\it bosonic} and {\it Wu} pictures, 
which all lead to the same physical results.
To further illustrate this fact, we plotted in Fig.~\ref{nFB} the populations
and the quasi-particle DOS in the fermionic and bosonic pictures, for
the same disordered system analyzed before ($\sigma_0=2$, $\Delta\sigma = 4$, $T=E_{\rm F}/\kB$).
As expected all the populations in fermionic description, $n^{(+)}_{\xi i}$,
are smaller than unity,
while some of the bosonic populations, $n^{(-)}_{\xi i}$, 
indicate a large increase at low energies.
This difference regarding the populations in the two pictures is compensated 
by the behavior of the quasi-particle DOS, $\tilde\sigma^{(+)}$ and 
$\tilde\sigma^{(-)}$, 
so that the particle densities $\rho_{\xi i}$ and $\rho_{\xi}$ are the same in any 
description. Specifically, we have for low energies 
$\tilde\sigma^{(-)}<\tilde\sigma^{(+)}$, while at high energies both $\tilde\sigma^{(-)}$ and $\tilde\sigma^{(+)}$ 
asymptotically approach the Wu density of states, $\tilde\sigma^W$.

An important aspect, which has consequences in the thermodynamical behavior
of the systems, is related to the reduction of available states especially 
in the sub-volumes with low DOS.
As it was pointed out in Section \ref{23sp}, this depletion occurs 
in the species which correspond to the lower quasi-energy spectrum and can lead to a complete spatial segregation of quasiparticles at a certain energy. 
In our case, the bottom energy
species in sub-volumes 10 and 18, followed by 6 and 19 have the lowest
occupation, as an effect of statistical (repulsive) interactions 
exerted by the neighboring species.
The depletion of energy levels has consequences not only in the
thermodynamic behavior, e.g. heat capacity, but also in the transport
properties, e.g. in systems with nearest neighbor hopping mechanism.

The heat capacity of such disordered systems exhibits peculiar effects. 
As it was indicated in Section \ref{23sp}, in a uniform system, 
i.e. with a constant DOS, the heat capacity 
is the same for any diagonal single FES parameter $\alpha$.
However, here we have FES parameters of the form (\ref{alphas_def1}),
which are diagonal in energy (since $d\sigma_\xi(\epsilon)/d\epsilon=0$ for any $\xi$) 
and non-diagonal with respect to position indices.
The results are presented in Fig.~\ref{Cv_T_20_50}(a) for different
distributions of the local DOS, as indicated in the inset, corresponding to
the values $\Delta\sigma = 1, 0.75, 0.5 \sigma_0$.
By increasing the disorder a deviation from the reference constant-DOS dependence
of the heat capacity per particle is observed, with a maximum above 1.
Similar deviations are observed in the temperature dendence of 
the entropy in Fig.~\ref{Cv_T_20_50}(b), 
which are in agreement with relation (\ref{CvS}). 
These deviations vanish in the low disorder or high temperature limits.
Ensemble averages on disorder were also performed and a qualitatively
similar behavior was found. 

The model presented here for interacting particles with disorder assumes an ergodic
behavior and the equilibrium thermodynamics is extracted for a single phase.
However from the dynamical perspective, systems of this type have typical
glassy behavior, which is generally quasi-ergodic and characterized 
by a sequence of relaxation time scales.
The FES parameters given by (\ref{alphas_def1}) 
can also provide the dynamics of the system following
the Monte Carlo approach described in Ref. \cite{JStatMech.P09011.2010.Nemnes}.

\section{Conclusions}

We have formulated an approach based on the fractional exclusion statistics (FES) to calculate the thermodynamic properties of a general class of systems of interacting particles.
The systems may have (in principle) any number of dimensions, may be either homogeneous or non-homogeneous, and we consider only particle-particle interactions which depend on the distance between the particles.
If the system exists in an $s$-dimensional space, then the species are defined in the $s+1$-dimensional space of positions and quasiparticle energies. 
This method allows the calculation of thermodynamic properties of relatively large particle systems, in a semi-classical fashion, yet incorporating the local properties of interacting quantum gases by means of statistical parameters.


For the consistency and for making connection with other approaches of FES, we presented three perspectives of the formalism called here {\it fermionic}, {\it bosonic} and {\it Wu}'s perspective.

To show the feasibility of our procedure, we applied it to test cases of different complexities, ranging from spatially homogeneous systems, to systems characterized by random distributions of local densities of states. 
We first analyzed test-case systems with two sub-volumes, where analytical calculations can be performed, and found a rather different thermodynamical behavior depending on the set of particle-particle interacting potentials.
In the framework of non-diagonal FES parameters, the chosen repulsive interactions can reduce the number of available states in the species with lower DOS, causing a quasiparticle population inversion. 
This has observable consequences in the temperature dependence of the heat capacity.

The FES formalism employed can describe the spatial accumulation and depletion of particles in systems with non-uniform DOS.
We considered here the screened Coulomb interaction, as it is of broad interest especially in the physics of semiconductor devices, where charging effects  at different interfaces are particularly important.
However, the formalism is not limited to this type of potentials and the FES parameters can be extracted for rather general many-body interactions. 

Prototypical glassy systems were investigated, where the real space disorder of sites was modeled by a random DOS.
Qualitative features found in the first test-cases analyzed, with two sub-volumes, are also present in the considered disordered systems. 
In particular, the repulsive interactions and the cooperative effect inherently set up by the finite range interactions leads to the depletion of states in the species with low quasiparticle energies, especially in sub-volumes with low DOS.
This causes a spatial segregation of quasiparticles at low energies, which implies consequences in the transport properties as well.
The obtained deviations in the temperature dependence of the heat capacity are proved to be proportional to the degree of disorder.

To conclude, the proposed model and method based on FES explores the thermodynamic behavior of non-homogeneous interacting-particle systems.
The feasibility of our method is proved by numerical calculations of thermodynamic properties of a variety of systems of different complexities. 
Elsewhere \cite{PhysRevE.88.042150.2013.Anghel} we compared the mathematical formalism of this approach with Landau's Fermi liquid theory and found that they are equivalent proving in this way the correctness of our formulation.
The quasiparticle analysis points out different thermodynamical features, which are correlated with the considered interacting potentials.
By including the spatial dimension, the presented FES formalism becomes also suitable for the investigation of transport properties of mesoscopic systems.

Another FES method, complementary to ours, has been recently proposed \cite{arXiv:1311.7574.Bakhti}. 
Our method applies to non-homogeneous systems with long range interactions in any number of dimensions or external potential (see also \cite{PhysRevE.88.042150.2013.Anghel}) if the number of particles and states within the range of the particle-particle interaction potential is large, whereas the method of Ref. \cite{arXiv:1311.7574.Bakhti} describes one-dimensional systems of equal free-particle energies (on-site energies), but for any particle-particle interaction range.


\section*{Acknowledgements}
%

The work was supported by the Romanian National Scientific Research Council projects PN-II-ID-PCE-2011-3-0960 and PN09370102/2009. The travel support from the Romania-JINR Dubna collaboration project Titeica-Markov is gratefully acknowledged.

\bibliographystyle{unsrt}

\end{document}